\documentclass[prb,aps,twocolumn,showpacs,floats]{revtex4}
\usepackage{graphicx}

\begin{document}

\title{Finite temperature semimetal-insulator transition on the honeycomb lattice}

\author{Minh-Tien Tran$^{1,2}$ and Kazuhiko Kuroki$^3$}
\affiliation{
$^1$Institute of Physics, Vietnamese Academy of Science and Technology, P.O.Box 429, 10000 Hanoi, Vietnam. \\
$^2$Asia Pacific Center for Theoretical Physics, POSTECH, Pohang, Gyeongbuk 790-784, Republic of Korea. \\
$^3$Department of Applied Physics and Chemistry, 
University of Electron Communication, Chofu, Tokyo 182-8585, Japan.}

\begin{abstract}
A semimetal-insulator transition in the Hubbard model
on the honeycomb lattice is studied by using the dynamical mean field theory. 
Electrons in the honeycomb lattice
resemble the Dirac electron liquid and for weak interactions the system is semimetal.
With increasing the local interaction a semimetal-insulator transition occurs.
We find a nonanalytical structure of the phase transition which consists of a first-order transition
line ending in a second-order transition point and high-temperature crossover line.
A phase separation of semimetal and insulator occurs at low temperatures. 
Maxwell construction is performed to determine the first order transition line. 
The phase diagram is also presented.  
\end{abstract}


\maketitle

\section{Introduction}
The study of correlation driven metal-insulator transition (MIT) has attracted great attention in recent
years. The MIT is realized in a number of transition metal oxides and organic compounds by application
of the pressure or chemical substitutions.\cite{Imada} Far from the transition point the metallic phase is 
well described by the Fermi liquid theory while in the insulating phase the electrons are localized.
When the magnetic frustration is large, the MIT occurs in the paramagnetic phase.
The metallic state is a Fermi liquid with a renormalized mass. 
The renormalized mass increases as the transition is approached. This is the essence of the Brinkman-Rice
theory of the MIT.\cite{Brinkman} The Brinkman-Rice scenario of the MIT is substantially developed 
by the dynamical
mean field theory (DMFT) of the Hubbard model in the paramagnetic phase.\cite{Metzner,Georges,GKKR} 
 The essential 
features of the MIT studied within the DMFT is a nonanalytical structure of the phase transition, which 
consists of the first-order transition line ending in a second-order transition point and the high temperature 
crossover line.\cite{GKKR,Krauth,Rozenberg} 
The nonanalytical structure of the MIT has been observed 
experimentally,\cite{Limelette} as well as has been confirmed by 
cluster DMFT\cite{Parcollet,Park} and by other techniques.\cite{Onoda}  

Recently, the experimental realization of a single layer of graphite, known as graphene,\cite{Novosenov,Neto}
has brought up renewed interest in the low temperature physics of the electrons on the honeycomb lattice. 
In the honeycomb lattice at half filling the noninteracting Fermi surface collapses into the edge points of
the Brillouin zone.
The tight-binding dispersion exhibits the Dirac cone near these points, and the density of states (DOS)
at the Fermi level vanishes. The electrons on the honeycomb lattice closely resemble the one of
massless Dirac fermions in $2+1$ dimensions. In particular, the Hubbard model
on the honeycomb lattice can be considered as an asymptotic infrared massive quantum electron dynamics 
in $2+1$ dimensions.\cite{Giuliani}
Therefore, the electrons on the honeycomb lattice
provide a condensed matter analogy of relativistic physics of electrons. 

The honeycomb lattice is also a basis structure of a number of materials such as 
magnesium diboride\cite{Akimitsu} or layered nitride superconductors\cite{Tou,Taguchi,Kitora}. 
Electron correlations
constitute the essential properties of these materials such as 
unconventional superconductivity.\cite{Baskaran,Baskaran1} The emergence of electron correlations
and the specific features of the honeycomb lattice structure underlies the material properties.
The Hubbard model on the honeycomb lattice is a minimal model to describe the emergence 
of electron correlations in the specific lattice structure.   

For weak local interactions the electrons on the honeycomb lattice always stay in the
paramagnetic state. There is no presence of superconducting or magnetic instabilities
at weak coupling.\cite{Giuliani} The interaction only renormalizes the Fermi velocity.
With increasing the local interaction the renormalized velocity decreases. When the renormalized
velocity vanishes the electrons are localized and the state is insulating.\cite{Jafari} It is a scenario of
the semimetal-insulator transition (SMIT) on the honeycomb lattice. The transition occurs in
the paramagnetic phase, and it is a version of the Mott MIT. However,
due to the absence of the 
quasiparticle mass, the Brinkman-Rice scenario of the MIT cannot be applied to the honeycomb lattice.
Parallel to the SMIT, at low temperatures the local interaction in the honeycomb lattice can lock electrons with
different spins into the different sublattices that creates a magnetic long-range order.
The numerical simulations also find a semimetal - antiferromagnetic insulator transition (SMAFIT) on the
honeycomb lattice at low temperatures.\cite{Sorella,Martelo,Pavia} This transition is a type of the
Slater MIT which is driven by long-range order. In the honeycomb lattice the Mott and the Slater MIT
compete with each other. However, when the magnetic frustration is strong they can destroy the magnetic
long-range order, and there is only the Mott transition. 
The previous DMFT studies of the SMIT 
on the honeycomb lattice showed that the SMIT is a second-order transition
in the paramagnetic phase,\cite{Jafari,Santoro} whereas the variational calculations showed a 
first-order transition characteristic of the SMIT.\cite{Martelo} 
In this paper we reexamine the SMIT on the honeycomb lattice by the DMFT. We
find a nonanalytical structure
of the SMIT on the honeycomb lattice which has not been pointed out in 
the previous DMFT studies.\cite{Jafari,Santoro} 
The nonanalytical structure is reminiscent to the phase
structure of the MIT in the square or the Bethe lattices where the DOS at the 
Fermi level is finite.\cite{GKKR,Krauth,Rozenberg} It also
consists of the first-order transition line ending in a second-order transition point and 
the high temperature crossover line. However, in contrast to the square or the Bethe lattices, 
the SMIT on the honeycomb lattice does not accompany the appearance and the disappearance
of a quasiparticle peak at the Fermi level. The absence of a quasiparticle state at the Fermi 
level is a specific feature of the electron dynamics in the honeycomb lattice. The SMIT on the
honeycomb lattice is accompanied with the appearance and the disappearance of a pseudogap structure
near the Fermi level. 

The outline of the present paper is as follows. In Sec.~II we describe the DMFT for
the Hubbard model in the honeycomb lattice. Numerical results are presented in Sec.~III.  
In Sec.~IV conclusion and remarks are presented.

\section{Model and method}
We consider the Hubbard model on the honeycomb lattice. 
The Hamiltonian of the system is
\begin{equation}
H = - t \sum_{<i,j>,\sigma} c^{\dagger}_{i\sigma} c^{\null}_{j\sigma} -
\mu \sum_{i} c^{\dagger}_{i\sigma} c^{\null}_{i\sigma} \nonumber + 
U \sum_{i} n_{i\uparrow} n_{i\downarrow} ,
\label{hubbard}
\end{equation} 
where $c^{\dagger}_{i\sigma}$ and $c_{i\sigma}$ are the creation and 
the annihilation operator of electrons at site $i$ 
with spin $\sigma$. $n_{i\sigma}=c^{\dagger}_{i\sigma} c_{i\sigma}$ is the density operator.
$t$ is the hopping integral between nearest neighbor
sites $i$ and $j$. We will take $t$ as the unit of energy. $U$ is the local interaction and $\mu$ is the chemical
potential.  
In the following we will consider only the 
half filling case, i.e. $\mu=U/2$.
The honeycomb lattice has a specific feature in the band structure, 
where the Fermi surface of noninteracting electrons at 
half filling is just the edge points of the Brillouin zone. The tight-binding dispersion near these
points exhibits the Dirac cone like the relativistic electrons and the DOS at the Fermi level
vanishes. The noninteracting electrons on the honeycomb lattice is a semimetal which has a zero gap
at the Fermi level. The specific feature distinguishes the honeycomb lattice from  other
lattices such as the square or the Bethe lattices where the DOS at the Fermi level is finite.
Apparently, in the honeycomb lattice the concept of effective mass is not appropriate to describe the electron
properties. As a consequence, the standard description of the Mott MIT is not valid in the honeycomb lattice.
To reveal the nature of the SMIT in the honeycomb lattice we use the DMFT.\cite{Metzner,Georges,GKKR} 
The DMFT is exact in infinite dimensions. However, for two dimensional lattices the DMFT neglects 
nonlocal correlations. The cluster DMFT studies for a square lattice have  shown that the key features of the
MIT are already captured by the single-site DMFT.\cite{Park} Thus, one can expect a similarity for
the honeycomb lattice. 

Since the honeycomb lattice is a Bravais lattice with a basis of two lattice sites, the
electron Green function can be written in the form of matrix $2\times 2$
\begin{equation}
\mathbf{G}(\mathbf{k},i\omega_n) =[\mathbf{G}_{0}^{-1}(\mathbf{k},i\omega_n) - 
\mathbf{\Sigma}(\mathbf{k},i\omega_n) ]^{-1} ,
\label{dyson} 
\end{equation} 
where $\omega_n=(2 n-1) \pi T$ is the Matsubara frequency, 
$\Sigma(\mathbf{k},i\omega_n)$ is the self energy, and $\mathbf{G}_{0}(\mathbf{k},i\omega_n)$ is
the bare Green function of noninteracting electrons. The bare Green function is
\begin{equation}
\mathbf{G}_0^{-1}(\mathbf{k},i\omega_n) =
\left( \begin{array}{cc}
i\omega_n +\mu & - \varepsilon(\mathbf{k}) \\
- \varepsilon^{*}(\mathbf{k}) & i\omega_n +\mu
\end{array} \right) ,
\end{equation} 
where $\varepsilon(\mathbf{k})=-2 t \exp(i k_x/2) \cos(\sqrt{3} k_y/2)- t \exp(i k_x)$. 
Equation (\ref{dyson}) is just the Dyson equation.
Within the DMFT,  the self energy is approximated by a local function of frequency, i.e.,
\begin{equation}
\mathbf{\Sigma}(\mathbf{k},i\omega_n) \approx \delta_{\alpha\beta} \Sigma(i\omega_n) .
\label{selfidmf}
\end{equation}
Note that within the DMFT the off diagonal elements of the self energy are neglected. 
These elements vanish in infinite dimensions. In finite dimension lattices they 
are indeed nonlocal correlation quantities.  
The self energy $\Sigma(i\omega_n)$ is determined from the dynamics of 
single-site electrons embedded in an effective mean field medium. Once the 
effective single-site problem  is solved the
self energy is calculated by the Dyson equation
\begin{equation}
\Sigma(i\omega_n)=\mathcal{G}^{-1}(i\omega_n) - G^{-1}(i\omega_n) ,
\label{dyson1}
\end{equation}
where ${\mathcal{G}}(\omega)$ is the bare Green function of the effective single site
and represents the effective mean field acting on the site. $G(\omega)$
is the electron Green function of the effective single site. The self consistent
condition requires that the Green function $G(\omega)$ of the effective single site
must coincide with
the local Green function of the original lattice. i.e.,
\begin{equation}
G(i\omega_n) = \frac{1}{N} \sum_{\mathbf{k}} G_{\alpha\alpha}(\mathbf{k},i\omega_n) ,
\label{cons}
\end{equation} 
where $N$ is the number of lattice sites.
Equations (\ref{dyson})-(\ref{cons}) form the self consistent system of equations
for the lattice Green function and the self energy. They are principal equations
of the DMFT. In order to solve the effective single site problem we use the exact diagonalization
technique.\cite{GKKR,Caffarel} 
The exact diagonalization maps the effective single site problem into an Anderson impurity
model
\begin{eqnarray}
H_{\rm{AIM}} &=& \sum_{p\sigma} E_p b^{\dagger}_{p\sigma} b^{\null}_{p\sigma} + 
\sum_{p\sigma} V_{p} (b^{\dagger}_{p\sigma} c_\sigma + \rm{h. c.} ) \nonumber \\
&& - 
\mu c^{\dagger}_\sigma c_\sigma + U n_{\uparrow} n_{\downarrow}  ,
\label{aim}
\end{eqnarray}
where the local impurity represented by $c^{\dagger}_\sigma$, $c_\sigma$ couples to a bath
of free conduction electrons represented by $b^{\dagger}_{p\sigma}$,  
$b^{\null}_{p\sigma}$ with dispersion $E_p$ via a hybridization $V_p$.
In the exact diagonalization the effective medium Green function $\mathcal{G}(i\omega_n)$ is
approximated by the corresponding Green function $\mathcal{G}_{n_s}(i\omega_n)$ calculated within
the Anderson impurity model of a finite number of bath levels
\begin{eqnarray}
\mathcal{G}_{n_s}(i\omega_n)^{-1} = i\omega_n + \mu - \sum_{p=2}^{n_s} \frac{|V_p|^2}{i\omega_n - E_p}, 
\end{eqnarray} 
where $n_{s}-1$ is the number of bath levels.
The model parameters $E_p$ , $V_p$ are determined by minimization of the distance function 
\begin{eqnarray}
d=\frac{1}{M} \sum_{n=1}^{M} |\omega_n|^{-k}|\mathcal{G}(i\omega_n)^{-1} - \mathcal{G}_{n_s}(i\omega_n)^{-1}|^2.
\end{eqnarray}
The parameter $k$ if chosen large ($k>1$) enhances the importance of the lowest Matsubara frequencies in the
minimization procedure. In particular, we take $k=3$ in the following numerical calculations. When the model
parameters $E_p$ , $V_p$ are obtained we solve the Anderson impurity model by the exact diagonalization and
obtain the local Green function $G(i\omega_n)$, and then the self energy $\Sigma(i\omega_n)$. Thus, we
obtain a closed self consistent system of equations for determining the electron Green function within the DMFT.

\section{Numerical results}

\begin{figure}[t]
\vspace{-0.5cm}
\includegraphics[width=0.48\textwidth]{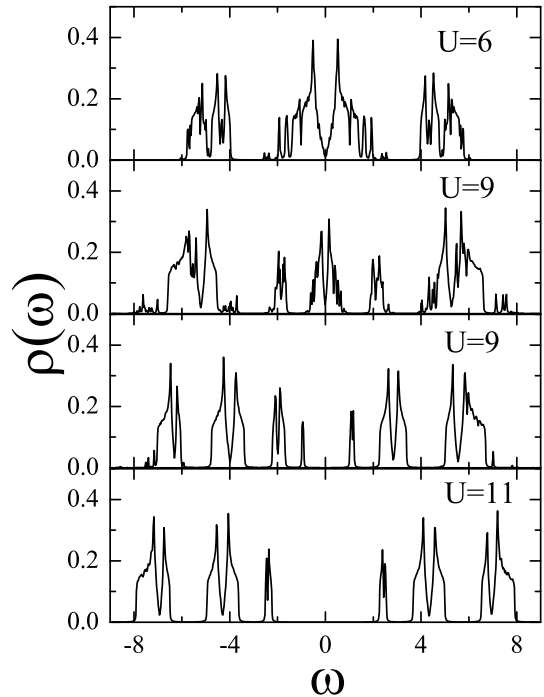}
\vspace{-0.5cm}
\caption{Density of states $\rho(\omega)$ for various $U$ at temperature $T=0.01$. 
The first two top panels plot the semimetal solution while the last two panels plot
the insulator solution.}
\label{figa}
\end{figure}

We solve the DMFT equations by iterations.\cite{GKKR,Caffarel} 
Most calculations are performed with $M=1024$ positive Matsubara
frequencies and lattice size of $64\times 64$ sites. For very low temperature (for instance $T=0.005$) we
take $M=2048$. The exact diagonalization of the Anderson impurity model is performed with $n_s=7$. We have
checked the agreement between $\mathcal{G}(i\omega_n)$ and $\mathcal{G}_{n_s}(i\omega_n)$ and found a good
agreement for whole model parameter range under consideration. 
We find two typical solutions, one is semimetal and the other is insulator. The 
DOS of these solutions is presented in Fig.~\ref{figa}.
The semimetal solution is characterized by a pseudogap near the Fermi 
level, while the insulator solution opens a wide gap at the Fermi level. 
In the semimetallic phase the DOS shows two Hubbard subbands and the pseudogap structure between them. 
In contrast to the square or the Bethe lattices, no Kondo quasiparticle peak appears at
the Fermi level. In the honeycomb lattice the DOS of noninteracting electrons linearly vanishes at
the Fermi level, so that the Kondo-singlet formation resulted in the effective single-site problem is
suppressed.\cite{Ingersent} The feature of the noninteracting DOS is retained in the interacting case,
so that the relativistic properties of the electrons in the honeycomb lattice are maintained as far as
the system does not approach the  SMIT. The local interaction only renormalizes the Fermi 
velocity which can be seen by the increase of the slope of the DOS near the Fermi level when 
the interaction increases. 
When the slope of the DOS near the Fermi level becomes very large, it closes the pseudogap
and the system transforms to the insulating phase. At the point of the SMIT the pseudogap
structure disappears and leaves a wide gap in the DOS. In the insulating phase the DOS exhibits only two
Hubbard subbands separated by the gap. The SMIT scenario in the honeycomb lattice is reminiscent to
the MIT in the square or the Bethe lattices. The crucial different feature is the absence
of the Kondo quasiparticle state at the Fermi level in the honeycomb lattice. As a consequence, the Brinkman-Rice
scenario of a divergence of the effective mass at the transition point is not valid. Instead, in
the honeycomb lattice the renormalized Fermi velocity vanishes when the system crosses the transition point.
This SMIT scenario was also observed in the DMFT studies with the iterated perturbation theory as the
impurity solver.\cite{Jafari} However, at finite temperature we observe a coexistence of
the semimetallic and the insulating solutions at intermediate interactions which has not been
pointed in the previous DMFT studies.\cite{Jafari,Santoro}    

\begin{figure}[t]
\vspace{-1cm}
\includegraphics[width=0.48\textwidth]{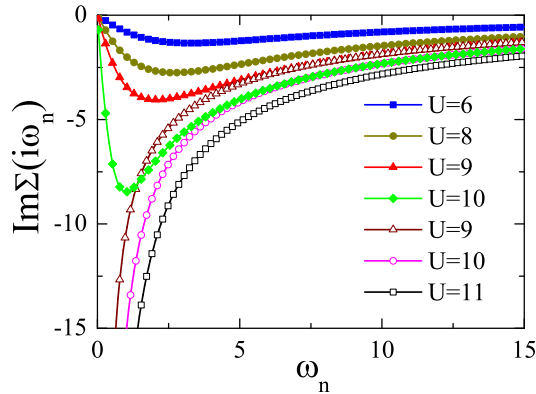}
\vspace{-1cm}
\caption{(Color online) The imaginary part of the self energy
for various interactions $U$ at temperature $T=0.01$. The filled symbols
are the self energy in the semimetallic phase, while the open symbols are
the self energy in the insulating phase.}
\label{figb}
\end{figure}

In Fig.~\ref{figb} we present the imaginary part of the self energy for various interactions in both
the semimetallic and the insulating phases. The slope of ${\rm Im}\Sigma(i\omega)$ for 
$\omega \rightarrow 0$ is identical to the slope of ${\rm Re}\Sigma(\omega+i 0^{+})$ for 
$\omega \rightarrow 0$. The renormalized factor of the Fermi velocity is
\begin{eqnarray} 
Z = \frac{\displaystyle 1}{ \displaystyle 1 -
\frac{\displaystyle \partial {\rm Re}\Sigma(\omega+i 0^{+})}{\partial \omega}
\bigg|_{\omega=0}} .
\label{Z}
\end{eqnarray} 
Figure \ref{figb} shows that in the semimetallic phase ${\rm Im}\Sigma(i\omega) \rightarrow 0$ as
$\omega \rightarrow 0$. It leads the DOS to be vanished at the Fermi level. As the interaction
increases the slope of ${\rm Im}\Sigma(i\omega)$ for 
$\omega \rightarrow 0$ increases, so that the renormalized factor $Z$ gradually decreases.
When the system transforms to the insulating phase the renormalized factor $Z$  vanishes.
Apparently, in the insulating phase the concept of the Fermi velocity is not valid and 
the use of Eq.~\ref{Z} does not make sense.  
Nevertheless, at the SMIT point the slope of ${\rm Im}\Sigma(i\omega)$ for 
$\omega \rightarrow 0$ abruptly changes from a negative large value to a positive large value. 
It is a specific feature
of the SMIT in the honeycomb lattice. In the square or the Bethe lattices, the slope of 
${\rm Im}\Sigma(i\omega)$ for $\omega \rightarrow 0$ continuously changes as the system crosses the
MIT.\cite{GKKR,Krauth,Rozenberg}  

\begin{figure}[b]
\vspace{-0.5cm}
\includegraphics[width=0.48\textwidth]{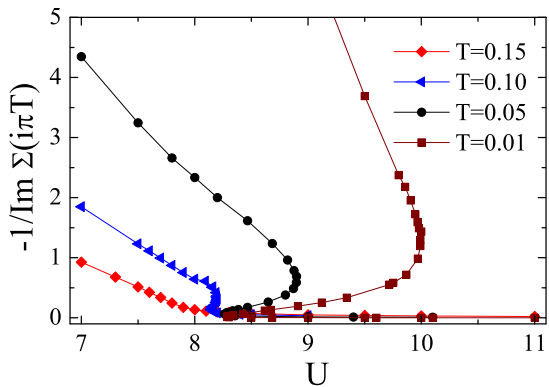}
\caption{(Color online) The quantity $Q=-1/\rm{Im}\Sigma(i \pi T)$ as a function
of $U$ at different temperature $T$.}
\label{fig1}
\end{figure}

\begin{figure}[b]
\vspace{-0.5cm}
\includegraphics[width=0.48\textwidth]{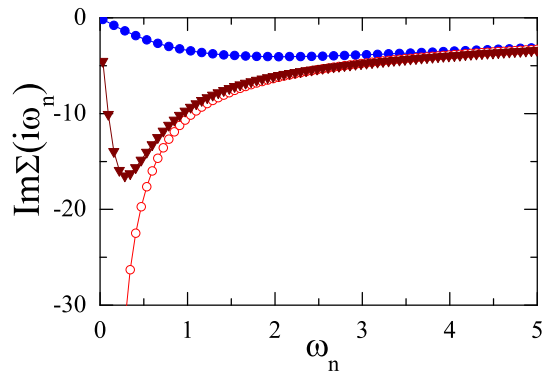}
\caption{(Color online) Three coexistent solutions at $U=9$ and $T=0.01$.
The imaginary part of the self energy is plotted. 
The semimetal solution is presented by the blue filled circles, the insulator solution
is presented by the red open circles, and the metastable solution is presented by
the brown filled triangles.  }
\label{fig2}
\end{figure}

At low temperatures we have found the insulator solution 
for $U>U_{c1}(T)$, and
the semimetal solution for $U<U_{c2}(T)$. In the range $U_{c1}(T) < U < U_{c2}(T)$ both insulator and
semimetal solutions coexist. The nonanalytical structure in the honeycomb lattice is very similar to the 
one in the square or the Bethe lattices.\cite{GKKR,Krauth,Rozenberg} This suggests that the SMIT in the honeycomb lattice
is a first order transition.  
In order to reveal the nonanalytical structure of the phase transition we use the method proposed
by Tong {\em et al}.\cite{Tong} It is based on the observation that for a fixed temperature
the formal dependence of a thermodynamical quantity $Q$ on the interaction $U$ is a multivalued
function $h_{Q}(U)$. Usually, $h_{Q}(U)$ has a "Z"- or "S"-shaped structure. The signal of the
nonanalytical structure is the discontinuity of $Q(U)$  in the normal calculations. To obtain the
multivalued function $h_{Q}(U)$, instead of $Q=h_{Q}(U)$ we transform it to a self-consistent
equation
\begin{equation}
Q = h_{Q}(U-\lambda [A - Q]) ,
\label{Q}
\end{equation} 
where $\lambda$ and $A$ are parameters which are chosen so that $Q$ is single valued with respect to $U$
even if the original $h_Q(U)$ is a multivalued function. Equation (\ref{Q}) is embedded into the DMFT
self-consistent equations. In the following calculations we take 
$Q=-1/\rm{Im}\Sigma(i \pi T)$ which is the inverse of the imaginary part of the self energy
at the first Matsubara frequency. This quantity is proportional to 
${\rm Im} \Sigma(i \omega_1)/ \omega_1 \approx \partial \rm{Re} \Sigma(\omega) / 
\partial \omega |_{\omega \rightarrow 0} $, which is a renormalized contribution to
the Fermi velocity. In Fig.~\ref{fig1} we present $Q$ as a function of $U$ at various temperatures.
One can see the "Z"-shaped structure of $h_Q(U)$ at low temperatures. At high temperatures $h_Q(U)$
is a single-value function. In the insulating phase $Q$  vanishes, while in
the semimetallic phase it is finite. The vanishing of $Q$ also indicates the vanishing of the
renormalized Fermi velocity.
At low temperatures in the range $U_{c1}(T) < U < U_{c2}(T)$ we find an 
additional solution to the semimetal and the insulator solutions.  An example of these three coexistent solutions are plotted
in Fig.~\ref{fig2}. One solution is semimetal (filled circles) and the other is insulator  
(open circles). The additional solution (filled triangle) is found between them. It fills the
positive slope piece of the multivalued function $h_Q(U)$. At high frequencies the additional
solution closely approaches to the insulator solution, while at low frequencies it behaves like
metallic. With the three coexistent solutions the function $h_Q(U)$ is continuous but multivalued
in the region $U_{c1}(T) < U < U_{c2}(T)$, as can be seen in Fig.~\ref{fig1}.  

\begin{figure}[t]
\vspace{-0.5cm}
\includegraphics[width=0.5\textwidth]{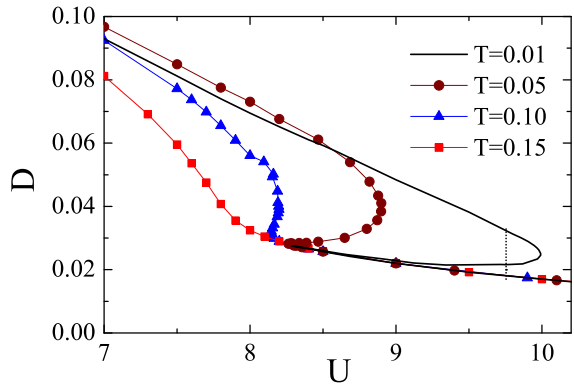}
\caption{(Color online) The double occupation $D$ as a function of $U$ at
various temperatures. The solid line, the filled cycles, triangles and squares are
the double occupation at temperature $T=0.01$, $0.05$, $0.1$, $0.15$, respectively.
The vertical dashed line is a Maxwell construction for $T=0.01$.}
\label{fig3}
\end{figure}

However, not all three solutions are stable. To find a stable solution we compare the free energies 
of the three coexistent solutions.
The free energy can be calculated via the double occupation\cite{Tong}
\begin{equation}
F(U,T)=F(0,T)+\int_{0}^{U} D(U',T) dU',
\end{equation}
where $F(U,T)$ is the free energy and $D(U,T)=\langle n_{\uparrow} n_{\downarrow} \rangle$ 
is the double occupation. The double occupation can be calculated by the exact 
relation\cite{Vilk}
\begin{equation}
D = \frac{T}{U} \sum_{n} G(i\omega_n) \Sigma(i\omega_n) e^{i\omega_n 0^{+}} . 
\end{equation}
The double occupation is a measure of the portion of lattice sites which are occupied
by electrons with both spins and characterizes the mobility degree of electrons in the lattice. 
For $U=0$, $D=0.25$ for $U=0$ and for $U \rightarrow \infty$, $D=0$. The double occupation
is often used to reveal the first order phase transition.\cite{GKKR,Tong} The DMFT shows that within a stable phase 
the double occupation decreases as U increases, and it exhibits a discontinuity when the system crosses a first order
transition line.\cite{GKKR,Tong} Indeed, $D=\partial F/\partial U$ and $\partial D / \partial U = \partial^2 F/\partial U^2$.
$\partial D / \partial U <0 $ shows a stability of phase. 
The variable pair $D-U$ is analogous to the inverse density and pressure in the conventional
liquid-gas transition theory.\cite{Tong,Castellani}    
In Fig.~\ref{fig3} we present the double occupation $D$ as a function of $U$ at various
temperatures. 
It shows that the double occupation in the insulating phase is
independent on temperature. It means that the thermal fluctuations do not affect the degree of the
electron mobility in the insulating phase, because the gap strongly prevents the mobility of electrons.   
At high temperatures $D(U)$ is a single-value function, while
at low temperatures $D(U)$ has "Z"-shaped structure like
the quantity $Q$. The obtained function $D(U)$ is very similar to the one in the square and Bethe lattices.\cite{Tong}
At high temperatures the double occupation $D(U)$ is smooth; thus,
the system only crosses from semimetal to insulator. For low temperatures,
at $U_{c1}$ the double occupation $D(U)$ is continuous, but its slope is discontinuous.
It means that the transition at $U_{c1}$ is a second-order transition. 
At $U_{c2}$ the double occupation is smooth; thus the system only crossovers from a semimetal
to a metallic-like phase. However,
comparing the free energy, one can see that
the additional metallic-like phase has highest free energy, and therefore it is
unstable. This feature is similar to the conventional liquid-gas transition, where the
liquid and the gas phases coexist.
The semimetallic phase is stable for $U<U_{c}(T)$, while the insulator one
is stable for $U>U_{c}(T)$.  $U_{c}(T)$ can
be found by a Maxwell construction $F_{SM}(U_c,T)=F_{I}(U_c,T)$. At low temperatures 
as $U$ crosses $U_{c}(T)$
from below, a stable semimetallic phase transforms into a stable insulating phase. This transition is
accompanied with a finite jump $\Delta D$ of the double occupation, that it is a first-order transition. 
The finite low-temperature
SMIT in the honeycomb lattice is of first order. 
When $T \rightarrow 0$,
one can expect that $U_c \rightarrow U_{c2}$, and at $U_c$ there is no  jump of the double
occupation. The zero temperature SMIT in the honeycomb lattice is of second order.\cite{Jafari} However, this
second-order phase transition is special. It  emerges from the metastable coexistent phases. 
At zero temperature near the phase transition a metastable
phase still exists.    

\begin{figure}[t]
\vspace{-0.5cm}
\includegraphics[width=0.48\textwidth]{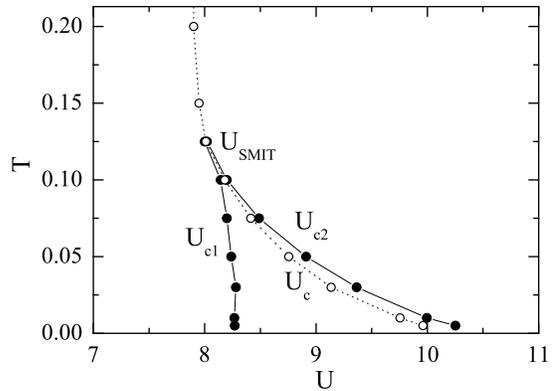}
\caption{Phase diagram of the SMIT in the honeycomb lattice.}
\label{fig4}
\end{figure}

We summarize the results in the phase diagram presented in Fig.~\ref{fig4}.
At low temperatures the semimetallic phase persists up to $U_{c2}(T)$, while the insulating
phase exists down to $U_{c1}(T)$.  Both lines $U_{c1}(T)$ and $U_{c2}(T)$ terminate at the point $U_{SMIT}$.
In the region $U_{c1}(T) < U < U_{c2}(T)$ three phases coexist. One phase is semimetallic, and the
other is insulating. The third phase is unstable. Actually, the SMIT transition occurs along the
line $U_c(T)$ which presents a first-order transition. The line $U_c(T)$ also terminates at
the second-order transition point $U_{SMIT}$. In the region $U_{c1}(T) < U < U_{c}(T)$ the semimetallic
phase is stable and the insulating phase is metastable, whereas in the region $U_{c}(T) < U < U_{c2}(T)$
the insulating phase is stable and the semimetallic phase is metastable. The phase separation
is the essential feature of the SMIT on the honeycomb lattice. This is reminiscent of the MIT in the square 
or the Bethe lattices. The feature may be considered as a common correlation effect 
regardless of the lattice structure. This may be unexpected because the MIT in the square or Bethe lattices
is accompanied with the appearance and disappearance of a quasiparticle peak which
is formed by the Kondo effect at the Fermi level. In the honeycomb lattice the Kondo effect is suspended and
the Kondo quasiparticle peak at the Fermi level is absent. The SMIT on the honeycomb lattice is accompanied
with appearance and disappearance of the pseudogap near the Fermi level.
Apparently, the phase separation in the MIT
is just an emergence of electron correlations in the boundary of metallic and insulating phases
without involving a specific transition driven mechanism. Near the zero temperature the first order phase transition
occurs at $U_c/t \approx 10$. The previous DMFT calculations adopting the iterated perturbation theory for single-site
problem obtained $U_c/t = 13.3$ at $T=0$.\cite{Jafari} It is well known that the iterated perturbation theory usually overestimates
the critical value.\cite{Bulla} The DMFT calculations for infinite dimension hyperdiamond 
lattice obtained $U_c/t \approx 8.5$.\cite{Santoro}
In graphene samples,\cite{Neto} $U/t \sim 2 \div 4$ is far from the SMIT. 
One may expect that the graphene is well described by the Dirac liquid theory with a 
renormalized velocity. 

\section{Conclusion}
In this paper we study the SMIT in the honeycomb lattice by using the DMFT. In contrast to the square
or Bethe lattices, the SMIT in the honeycomb lattice occurs without involving the appearance and
the disappearance of a quasiparticle state at the Fermi level. It is accompanied by the appearance
and the disappearance of a pseudogap near the Fermi level.
Far from the transition point the semimetallic phase is 
a Dirac electron liquid with a renormalized Fermi velocity, while in the insulating phase 
the electrons are localized. When the system approaches the SMIT point from below the renormalized
Fermi velocity vanishes. We found a nonanalytical
structure of the phase transition. It consists of a first-order transition
line ending in a second-order transition point and high-temperature crossover line. At low temperatures
the phase separation between the semimetallic and the insulating phases occurs. It suggests that the
phase separation in a common feature of the Mott MIT regardless of a specific transition driven mechanism.  
In two-dimensional lattices the DMFT neglects nonlocal correlations. The cluster DMFT calculations
for a square lattice shows that the first order characteristic of the Mott MIT are already captured in the single-site
DMFT.\cite{Park} However, the nonlocal correlations modify the shape of the transition lines.\cite{Park}
One may expect the same feature in the honeycomb lattice. Moreover, in the honeycomb lattice
the magnetic instability may compete with the Mott MIT at low temperatures. 
It requires a further study, at least, how strong short range magnetic correlations affect the Mott MIT. 

\begin{acknowledgments}

One of the authors (M.-T.) acknowledges the Nishina Memorial Foundation for the Nishina fellowship.

\end{acknowledgments}

\end{document}